\documentclass[aps,prr,twocolumn,showpacs,superscriptaddress,longbibliography]{revtex4-1}
\usepackage{graphicx}
\usepackage{amsmath}
\usepackage{amssymb}
\usepackage{hyperref}
\usepackage{siunitx}
\usepackage{algpseudocode}
\usepackage{algorithm}
\usepackage{mathtools}
\usepackage{bm}
\usepackage{comment}
\usepackage{xcolor}
\newcommand{\hitomi}[1]{\textcolor{black}{#1}}

\begin{document}

\title{Typical reconstruction limit and phase transition of maximum entropy method}

\author{Masaru Hitomi}
\affiliation{Graduate School of Information Sciences, Tohoku University, Sendai 980-8579, Japan}

\author{Masayuki Ohzeki}
\affiliation{Graduate School of Information Sciences, Tohoku University, Sendai 980-8579, Japan}
\affiliation{Department of Physics, Institute of Science Tokyo, Tokyo 152-8551, Japan}
\affiliation{Sigma-i Co., Ltd., Tokyo 108-0075, Japan}

\date{\today}

\begin{abstract}
    We investigate the dependence of the maximum entropy method (MEM) reconstruction performance on the default model. 
    The maximum entropy method is a reconstruction technique that utilizes prior information, referred to as the default model, 
    to recover original signals from observed data, 
    and it is widely used in underdetermined systems. 
    The broad applications have been reported in fields ranging from the analysis of observational data in seismology and astronomy, 
    to large-scale computations in quantum chemistry, and even in social sciences such as linguistics.
    However, a known drawback of MEM is that its results depend on the assumed default model.
    In this study, 
    we employ the replica method to elucidate how discrepancies in the default model affect the reconstruction of signals with specific distributions. 
    We report that 
    \hitomi{
        for binary signals
    }, 
    even small discrepancies can induce phase transitions, leading to reconstruction failure. 
    Additionally, 
    by comparing MEM with reconstruction based on $L_1$-norm regularization, a method proposed in recent years, 
    we demonstrate that MEM exhibits lower reconstruction accuracy under certain conditions.
\end{abstract}

\maketitle

\section{Introduction}
Maximum Entropy Method (MEM) is one of the techniques used to reconstruct original signals from limited and noisy observations \cite{Burg1967,Burg1968,burg1975maximum}. 
By assuming prior information, known as the default model and maximizing the entropy relative to this model, 
MEM reconstructs the original signal. 
Since the introduction as a method for seismological analysis~\cite{Burg1967,Burg1968,burg1975maximum}, 
MEM has been applied across a wide range of scientific disciplines, 
including 
geology
\cite{Burg1972,andersenCALCULATIONFILTERCOEFFICIENTS1974,radoskiComparisonPowerSpectral1975,jacksonMaximumEntropyRegularization2007,zoukaneriCombinedWignerVilleMaximum2015,liApplicationMaximumEntropy2019,wangMultichannelMaximumentropyMethod2020,martiniMaximumEntropySpectral2024}, 
quantum chemical calculations
\cite{silverMaximumentropyMethodAnalytic1990,gubernatisQuantumMonteCarlo1991,vonderlindenUnbiasedAccessExchange1993,bergeronAlgorithmsOptimizedMaximum2016,suiQuasiparticleDensityStates2016,krabergerMaximumEntropyFormalism2017,ghanemGeneralizedMaximumEntropy2023}, 
image processing
\cite{werneckeMaximumEntropyImage1977a,gull1978image,gull1984maximum,martin1985maximum,bazulin2018maximum},
economics
\cite{benedetto2015maximum},
and linguistics
\cite{kawahara2019accounting,shih2019gradient}. 
More recently, 
it has been utilized as one of the analytical techniques by the Event Horizon Telescope Collaboration (EHTC) to estimate the structure of the black hole shadow and jets observed in M87
\cite{Bouman2016,Akiyama2017,EHT2019I,EHT2019II,EHT2019III,EHT2019IV,EHT2019V,EHT2019VI,EHT2021VII,EHT2021VIII,EHT2023IX}.

In contrast, 
a newer reconstruction method, known as $L_1$-norm regularization has been proposed
\cite{Donoho2006,CandesWakin2008,DonohoTanner2005,CandesTao2005,CandesTao2006,Candes2008,Angelosante2009,Ye2019,KabashimaWadayamaTanaka2009,Krzakala2012,Keiper2017,doi2024phase}. 
This method assumes the original signal is sparse, eliminating the need for a default model required in MEM.
In practice, 
both MEM and $L_1$-norm regularization have been employed as part of the analyses in the EHTC, 
with each method producing reliable results
\cite{Bouman2016,Akiyama2017,EHT2019III,EHT2019IV}.

Given the inherently underdetermined nature of such reconstruction problems, 
ensuring the accuracy of these methods is critical importance. 
For $L_1$-norm regularization, 
statistical mechanics approaches such as the replica method have been employed to evaluate its typical reconstruction performance
\cite{KabashimaWadayamaTanaka2009,Krzakala2012,doi2024phase}. 
It has been reported that for certain problems, phase transitions between successful and unsuccessful reconstruction phases exist, 
highlighting the importance of understanding these phenomena for specific cases
\cite{Keiper2017,doi2024phase}.
On the other hand, 
despite the well-known dependency of MEM on the default model, systematic analyses of its reconstruction accuracy remain scarce
\cite{saitoPossibleInstabilityBurg1978}. 
If the assumed default model deviates from the true distribution of the original signal, 
the reconstructed results might be biased toward the default model. 
Although MEM is implicitly used in many fields as an analytical tool or library, 
the reliability of its outcomes is not guaranteed.

In this paper, 
we perform a typical performance evaluation of MEM using the replica method. 
By assuming simple binary signals as the original signals and introducing deviations in the default model, 
we investigate the conditions under which reconstruction succeeds or fails. 
We report that the reconstruction accuracy strongly depends on the deviation of the default model and that reconstruction failure can occur even with small deviations. 
Moreover, 
we identify a phase transition between success and failure phases, 
which depends on the sparsity of the original signal.

Additionally, 
we compare MEM with $L_1$-norm regularization for the same problem. 
Our findings reveal that MEM exhibits low reconstruction accuracy in both sparse and dense regions, 
despite $L_1$-norm regularization excelling in sparsity. 
This suggests that $L_1$-norm regularization often outperforms MEM, 
regardless of the sparsity of the original signal.

This paper is organized as follows. 
In Sec.~\ref{sec_method}, 
we present the problem settings and analytical methods. 
Specifically, 
we define the underdetermined problem for binary original signals, formulate the Maximum Entropy Method (MEM), 
and describe two distinct default models: the deviation model and the flipping model.
We also provide an explanation of the replica method as the analytical approach and discuss the numerical method based on the Alternating Direction Method of Multipliers (ADMM). 
Next,
we discuss the computational results obtained using each default model and analytical approach. 
In particular, 
we elucidate the existence of phase transition points, model dependence, and the breaking of symmetry related to sparsity in Sec.~\ref{sec_res}. 
As Sec.~\ref{sec_res_L1}, 
we compare the reconstruction results between $L_1$-norm regularization and MEM assuming the flipping model. 
Finally, Sec.~\ref{sec_con} provides the conclusion.

\section{Method} \label{sec_method}
\subsection{Problem settings} \label{sub_prob}
To evaluate the typical performance of the MEM, 
we consider the problem of estimating an unknown original signal $x$ from known measurements $y$ and $A$. 
The problem follows the linear equations,
\begin{equation}
    \bm{y} = A \bm{x}^0,
    \label{eq:linear_equation}
\end{equation}
where $\bm{x}^0 \in \mathbb{R}^{N}$ and $\bm{y} \in \mathbb{R}^{M}$ denote the original and observed signals,
and $A \in \mathbb{R}^{M \times N}$ is measurement matrix, respectively.
When $M < N$, the problem becomes underdetermined.

We assume that each element of the original signal $x$ is a binary variable $(0, 1)$, and its probability distribution is
\begin{equation}
    P(x^0_i) = \rho \delta(x^0_i - 1) + (1 - \rho) \delta(x^0_i),
    \label{eq:x0_distribution}
\end{equation}
where $\rho$ represents the proportion of nonzero components in $x$. 
Furthermore, each element of the observation matrix $A$ is assumed to follow the probability distribution,
\begin{equation}
    P(A_{ij}) = \mathcal{N} \left(0, \frac{1}{N} \right),
    \label{eq:A_distribution}
\end{equation}

On the base of this setup, 
we perform estimation using MEM. 
The solved problem is formulated as the following minimization problem with respect to relative entropy,
\begin{equation}
    \underset{\{0 \leq x \leq 1\}^N}{\min} -S(\bm{x}; \bm{w}) \quad \text{subject to} \quad \bm{y} = A \bm{x},
    \label{eq:S_minimization}
\end{equation}
where the entropy $S(\bm{x}, \bm{w})$ is defined as
\hitomi{
\begin{equation}
    -S(\bm{x}; \bm{w}) = \sum_{i=1}^{N}{\left( -x_i + w_i + x_i \log \frac{x_i}{w_i} \right)}
    \label{eq:entropy}
\end{equation}
}
\hitomi{
where $\bm{w} \in \mathbb{R}^{N}$ is a default model depending on the models discussed in next sections and
we adopt the convention that $0 \log 0 = 0$ in this work.
}

We note that we define the estimation values here as a relaxed problem with continuous values
for reconstruction using MEM,
though the original signal takes only discrete binary values.

\subsection{Deviation model} \label{sub_model1}
In the following two sections, we describe two different default models: the deviation model and the flipping model.
\hitomi{
    It should be noted that the deviation discussed here does not correspond to noise in the observed signal, 
    but rather characterizes the manner and extent to which the default model differs from the original signal. 
    In this work, we do not consider noise in the observed signal.
}

The default model of the deviation model is defined by the following probability distribution,
\hitomi{
\begin{equation}
    P_{\varepsilon}(w_i | x^0_i) = \delta(x^0_i - 1)\delta(w_i - 1) + \delta(x^0_i)\delta(w_i - \varepsilon),
    \label{eq:der_distribution}
\end{equation}
}
where $\varepsilon$ is a parameter representing the "deviation" of the default model when the element of the original signal $x_i = 0$. 
In this model, 
when $x_i = 1$, the corresponding default model element is set as $w_i = 1$.
On the other hand, 
when $x_i = 0$, the default model element is set as $w_i = \varepsilon$, 
introducing a deviation of $\varepsilon$ from the true value.


\subsection{Flipping model} \label{sub_model2}
The probability distribution of the default model in the flipping model is expressed as
\hitomi{
\begin{align}
    P_{\eta}(w_i | x^0_i) = \eta \delta(x^0_i + w_i - 1) + (1-\eta)\delta(x^0_i - w_i),
    \label{eq:flp_distribution}
\end{align}
}
where $\eta$ represents the probability that the element of the default model is flipped relative to the element of the original signal. 
More specifically,
when the element of the original signal is $x_i = 1$ ($0$), 
the value of default model is $w_i = 0$ ($1$) with probability $\eta$.
While the default model in the deviation model represents a continuous deviation from the true value, 
the flipping model evaluates a discrete deviation flipped from the true value. 
Therefore, both the original signal and the default model take discrete values.


\subsection{Replica analysis} \label{sub_rep}
To evaluate the typical performance of MEM, 
it is necessary to solve the minimization problem given by Eq.~\eqref{eq:S_minimization} for the given $\bm{y}$ and $A$,
and compare the results with the true value $\bm{x}^0$. 
However, solving this problem analytically is difficult. 
To avoid this difficulty,
we regard the constrained minimization problem given by Eq.~\eqref{eq:S_minimization} as a posterior distribution of the inverse temperature $\beta$,
thus ;
\begin{align}
    p(\bm{x}|A, \bm{y}) 
    &= \frac{p(A, \bm{y}| \bm{x})p(\bm{x})}{Z_{\beta}(A, \bm{y})} \nonumber \\
    &= 
    \frac{\delta(\bm{y} - A \bm{x})\exp(\beta S(\bm{x}; \bm{w}))}
    {\int_{0}^{1} d\bm{x}\, \delta(\bm{y} - A \bm{x})\exp(\beta S(\bm{x}; \bm{w}))}, 
    \label{eq:posterior}
\end{align}
where $Z_{\beta}(A, \bm{y})$ plays a role of a partition function.
By considering the limit $\beta \rightarrow 0$,
Eq.~\eqref{eq:posterior} generally converges to a uniform distribution over the solution of Eq.~\eqref{eq:S_minimization}.
\hitomi{
    When we consider
} 
the limit $\beta \rightarrow \infty$, 
the minimum value of $-S(\bm{x})$ under the given constraints can be determined. 
Regarding $\beta$ as the inverse temperature, 
this problem is equivalent to calculating the free energy $F$ at zero temprature in statistical mechanics. 
\hitomi{
    Here we note that the subsequent analysis is performed with the compression ratio $\alpha = M/N$,
    the non-zero elements rate in the original signal $\rho$ fixed, 
    considering the limit $N \rightarrow \infty$.
}
Furthermore, by assuming self-averaging on the free energy density $f$, 
it is possible to apply the replica analysis.
The free energy density is expressed as
\begin{align}
    f &= - \lim_{\beta \rightarrow \infty} \lim_{N \rightarrow \infty} \frac{1}{\beta N}
        \left[ 
            \ln Z_{\beta}(A, \bm{y})
        \right]_{A, \bm{x}^{0}, \bm{w}} \nonumber \\
    &= - \lim_{\beta \rightarrow \infty}
      \lim_{n \rightarrow 0}
      \frac{\partial}{\partial n}
      \lim_{N \rightarrow \infty} \frac{1}{\beta N}    
            \ln \left[ Z_{\beta}^{n}(A, \bm{y})
      \right]_{A, \bm{x}^{0}, \bm{w}} ,
      \label{eq:free_energy_density}
\end{align}
where $[~]_{A, \bm{x}^{0}}$ represents the configurational average with respect to $A$ and $\bm{x}^{0}$.
Assuming $n$ is a positive integer, 
we can express the expectation of $Z_{\beta}^{n}(A, \bm{y})$ as a replicated system.
We assess $Z_{\beta}^{n}(A, \bm{y})$ in Eq.\eqref{eq:free_energy_density} for $n \in \mathbb{Z}$, then the free energy density is evaluated 
by performing an analytic continuation for $n \in \mathbb{R}$. 
The following equation explicitly expresses the replicated system
\begin{align}
    &\left[Z_{\beta}^{n}(A, \bm{y})\right]_{A, \bm{x}^{0}, \bm{w}} \nonumber \\
    &= \left[
        \prod_{a=1}^{n}
        \int_{0}^{1} d \bm{x}^{a} 
        \prod_{\mu=1}^{M}
        \delta(y_{\mu} - \bm{a}_{\mu}^{T} \bm{x}^{a}) \exp(\beta S(\bm{x}, \bm{w}))
    \right]_{A, \bm{x}^{0}, \bm{w}} ,
    \label{eq:replicated_system}
\end{align}
where $\bm{x}^{a}$ is the vector for the $a$-th replica, 
$y_{\mu}$ denotes the $\mu$-th element of the observed signal $\bm{y} = A \bm{x}^{0}$, 
and $\bm{a}_{\mu}^{T}$ stands for the $\mu$-th row of the observation matrix $A$.
To perform calculation, we define the value
\begin{align}
    u_{\mu}^{a} = y_{\mu} - \bm{a}_{\mu}^{T} \bm{x}^{a} 
                = \bm{a}_{\mu}^{T}(\bm{x}^{0} - \bm{x}^{a}) .
    \label{eq:u_transformed}
\end{align}
Since each element of the observation matrix follows the Gaussian distribution $\mathcal{N}(0, 1/N)$, 
the expected value and covariance for $u_{\mu}^{a}$ become
\begin{align}
    \mathbb{E}\left[u_{\mu}^{a} \right]_{A} &= 0\\
    \text{Cov} \left[u_{\mu}^{a}, u_{\mu}^{b} \right]_{A}
    &= \frac{1}{N}{\bm{x}^{0}}^{T} \bm{x}^{0} 
    - \frac{2}{N}{\bm{x}^{0}}^{T} \bm{x}^{a} 
    + \frac{1}{N}{\bm{x}^{a}}^{T} \bm{x}^{b}.
    \label{eq:u_ave}
\end{align}

Here, we define the order parameters under the replica symmetric (RS) assumption.
\begin{align}
    & \rho = \frac{1}{N}\sum_{i=1}^{N}x_{i}^{0} = \frac{1}{N}{\bm{x}^{0}}^{T} \bm{x}^{0} ~~ (\because x_{i}^{0} \in \{0, 1\}) \label{eq:orders_rho} \\
    & m = \frac{1}{N}{\bm{x}^{0}}^{T} \bm{x}^{a} ~~ (a = 1 \dots n) \label{eq:orders_m} \\
    & Q = \frac{1}{N}{\bm{x}^{a}}^{T} \bm{x}^{a} ~~ (a = 1 \dots n) \label{eq:orders_Q} \\
    & q = \frac{1}{N}{\bm{x}^{a}}^{T} \bm{x}^{b} ~~ (a,b = 1 \dots n, \, a \neq b). \label{eq:orders_q}
\end{align}
It is capable to rewrite $u_{\mu}^{a}$ with the order parameters defined Eqs.~\eqref{eq:orders_rho}$\sim$\eqref{eq:orders_q} 
and the random variables $\xi_{a}$ and $z$ sampled Gaussian distribution $\mathcal{N}(0, 1)$ as follows:
\begin{align}
    u_{\mu}^{a} = \sqrt{Q - q} \, \xi_{a} + \sqrt{\rho - 2m + q} \, z.
\end{align}
The free energy density $f$ in the limit as $\beta \rightarrow \infty$, under the RS assumption is
\begin{widetext}
\begin{equation}
    f = \underset{{Q, m, \chi, \tilde{Q}, \tilde{m}, \tilde{\chi}}}{\text{extr}}
    \left\{
    - \frac{1}{2}Q\tilde{Q} + m\tilde{m} + \frac{1}{2}\chi \tilde{\chi}
    + \frac{\alpha}{2\chi}(\rho - 2m + Q)
    + \int Dt ~ \left[ \Psi (t,x^{0},w; \tilde{Q}, \tilde{m}, \tilde{\chi}, \tilde{p}) \right]_{x^{0},w}
    \right\} ,
    \label{eq:extr_free_energy_density}
\end{equation}
\end{widetext}
where
\begin{align}
    & \Psi (t,x^{0},w; \tilde{Q},\tilde{m},\tilde{\chi}) \nonumber \\
    &= \underset{\{0 \leq x \leq 1\}}{\text{min}} 
    \left\{ \frac{\tilde{Q}}{2} x^2 - (mx^0 + \sqrt{\tilde{\chi}}t)x - s(x; w) \right\},
    \label{eq:psi}
\end{align}
and $s(x; w)$ is the entropy element, 
\begin{equation}
    -s(x; w) = -x + w + x \log \frac{x}{w}.
\end{equation}
We can derive the saddle-point equations by extremizing Eq.~\eqref{eq:extr_free_energy_density} 
for the deviation model with the default model described in Eq.~:\eqref{eq:der_distribution},
\begin{align}
    & \tilde{Q} = \frac{\alpha}{\chi}, \label{eq:extr_der_Qtilde} \\
    & \tilde{m} = \frac{\alpha}{\chi}, \label{eq:extr_der_mtilde} \\
    & \tilde{\chi} = \frac{\alpha}{\chi^{2}}(\rho - 2m + Q), \label{eq:extr_der_chitilde} \\
    & Q = \int Dt ~ (\rho x^{*^2}(1,1) + (1-\rho) x^{*^2}(0,\varepsilon)), \label{eq:extr_der_Q} \\
    & m = \int Dt ~ \rho x^{*}(1,1), \label{eq:extr_der_m} \\
    & \chi = \int Dt ~ \left( \rho \frac{x^{*}(1,1)}{\tilde{Q}x^{*}(1,1)+1} + (1-\rho) \frac{x^{*}(0,\varepsilon)}{\tilde{Q}x^{*}(0,\varepsilon)+1} \right), \label{eq:extr_der_chi}
\end{align}
where $x^*(x^0,w)$ is the $x$ realizing minimization of Eq.~\eqref{eq:psi},
\begin{equation}
    x^*(x^0,w) = \max \left[0,~\min \left\{1,~\frac{1}{\tilde{Q}} W \left(w \tilde{Q} \mathrm{e}^{(mx^0 + \sqrt{\tilde{\chi}}t)} \right) \right\} \right],
    \label{eq:x_ast}
\end{equation}
and $W(z)$ denots Lambert's W function.
The equations for the flipping model (Eq.~:\eqref{eq:flp_distribution}) are given by
Eqs.~\eqref{eq:extr_der_Qtilde} $\sim$ \eqref{eq:extr_der_chitilde} and
\begin{align}
    Q &= \int Dt ~ \bm{\rho}^T X_Q \bm{\eta}, \label{eq:extr_flp_Q} \\
    m &= \int Dt ~ \bm{\rho}^T X_m \bm{\eta}, \label{eq:extr_flp_m} \\
    \chi &= \int Dt ~ \bm{\rho}^T X_{\chi} \bm{\eta}, \label{eq:extr_flp_chi}
\end{align}
where $X_Q$, $X_m$ and $X_{\chi}$ are the matrices,
\begin{align}
    X_Q = 
    \begin{pmatrix}
        x^{*^2}(1,0) & x^{*^2}(1,1) \\
        x^{*^2}(0,1) & x^{*^2}(0,0)
    \end{pmatrix}, \\
    X_m = 
    \begin{pmatrix}
        x^{*}(1,0) & x^{*}(1,1) \\
        0 & 0
    \end{pmatrix}, \\
    X_{\chi} = 
    \begin{pmatrix}
        \frac{x^{*}(1,0)}{\tilde{Q}x^{*}(1,0)+1} & \frac{x^{*}(1,1)}{\tilde{Q}x^{*}(1,1)+1} \\
        \frac{x^{*}(0,1)}{\tilde{Q}x^{*}(0,1)+1} & \frac{x^{*}(0,0)}{\tilde{Q}x^{*}(0,0)+1}
    \end{pmatrix}, \label{eq:X_chi}
\end{align}
and $\bm{\rho}$ and $\bm{\eta}$ are denoted as
\begin{align}
    \bm{\rho} =
    \begin{pmatrix}
        \rho \\
        1 - \rho
    \end{pmatrix},~~
    \bm{\eta} =
    \begin{pmatrix}
        \eta \\
        1 - \eta
    \end{pmatrix}.
\end{align}
The detail deviation of Eqs.~\eqref{eq:free_energy_density} $\sim$ \eqref{eq:X_chi} is represented in Appendix.
We also obtain an analytical expression for the typical mean squared error (MSE) value,
\begin{align}
    \text{MSE} = \rho - 2m + Q. \label{eq:MSE_replica}
\end{align}

\subsection{Numerical calculation} \label{sub_num}
To compare with the performance evaluation using the replica method, 
we propose a numerical reconstruction method based on the alternating direction method of multipliers (ADMM). 
It should be noted that the accuracy of reconstruction using ADMM depends on the tunable parameter 
and is applicable only to problems of finite size. 
Therefore, it does not necessarily coincide with the infinite size limit and typical performance assumed in the replica method.

We consider the minimization problem,
\begin{equation}
    \underset{\bm{x}}{\min} \left\{f(\bm{x}) + g(\bm{x}) \right\}.
    \label{eq:ADMM_min}
\end{equation}
Acording to the general representation of ADMM,
Eq.~\eqref{eq:ADMM_min} is regarded as a constrained problem
\begin{equation}
    \underset{\bm{x},\bm{z}}{\min} \left\{f(\bm{z}) + g(\bm{x}) \right\}~~s.t.~~\bm{x}-\bm{z}=0
    \label{eq:ADMM_min_const}
\end{equation}
and applying the augmented Lagrangian method, we obtain the cost function,
\begin{align}
    & L(\bm{x},\bm{z},\bm{h}[t]) \nonumber \\
    &= f(\bm{z}) + g(\bm{x}) + \left(\bm{h}[t]\right)^T (\bm{x}-\bm{z}) + \frac{\mu}{2} |\bm{x}-\bm{z}|^2,
    \label{eq:ADMM_min_cost}
\end{align}
where $\bm{h}[t]$ denotes the Lagrange multipliers and $\mu$ is the penalty parameter.
Considering the zero points of $L(\bm{x},\bm{z},\bm{h}[t])$, the update law is as follows;
\begin{align}
    \bm{x}[t+1] &= \underset{\bm{x}}{\text{argmin}} \left\{ L(\bm{x},\bm{z}[t],\bm{h}[t]) \right\}, \label{eq:ADMM_update_x} \\
    \bm{z}[t+1] &= \underset{\bm{z}}{\text{argmin}} \left\{ L(\bm{x}[t],\bm{z},\bm{h}[t]) \right\}, \label{eq:ADMM_update_z} \\
    \bm{h}[t+1] &= \bm{h}[t] + \mu (\bm{x} - \bm{z}). \label{eq:ADMM_update_h}
\end{align}
By substituting $f(\bm{z}) = -S(\bm{z}, \bm{w})$ and $g(\bm{x}) = |\bm{y} - A \bm{x}|$, 
we obtain the update law of MEM corresponding to Eqs.~\eqref{eq:ADMM_update_x},\eqref{eq:ADMM_update_z} as
\begin{align}
    \bm{x}[t+1] &= \underset{\bm{x}}{\text{argmin}} \{ \frac{1}{2\lambda} |\bm{y} - A \bm{x}|^2_2 \nonumber \\
    &+ (\bm{h}[t])^{T} (\bm{x}-\bm{z}[t]) + \frac{\mu}{2} |\bm{x}-\bm{z}[t]|^2_2 \}, \label{eq:ADMM_update_sp_x}
\end{align}
\begin{align}
    \bm{z}[t+1] &= \underset{\bm{z}}{\text{argmin}} \{ -S(\bm{z}, \bm{w}) \nonumber \\
    &+ (\bm{h}[t])^{T} (\bm{x}[t+1]-\bm{z})  + \frac{\mu}{2} |\bm{x}[t+1]-\bm{z}|^2_2 \}. \label{eq:ADMM_update_sp_z}
\end{align}
The numerical experiment is discussed in Sec.~\ref{sec_res}.

\section{Phase diagram} \label{sec_res}
\subsection{Deviation model} \label{sub_res_model1}
\begin{figure}[t]
    \centering
    \includegraphics[width=1.0\linewidth]{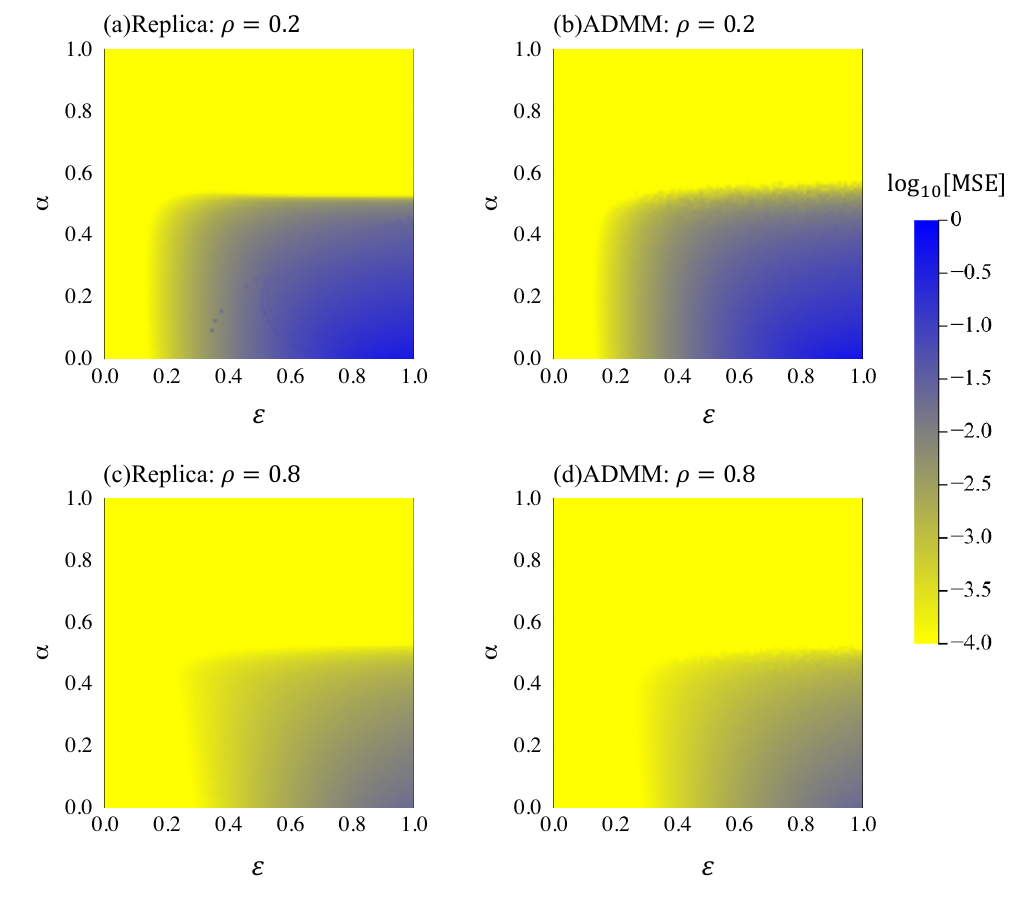}
    \caption{
        The phase diagram of the deviation model. 
        (a)The result of the replica analysis and (b)one of ADMM calculation for $\rho=0.2$.
        (c)The result of the replica analysis and (d)one of ADMM calculation for $\rho=0.8$.
    }
    \label{fig:phase_deviation}
\end{figure}

Figures~\ref{fig:phase_deviation}(a) and (c) show the phase diagram of the deviation model at $\rho = 0.2,~0.8$
calculated by the saddle equations, Eqs.~\eqref{eq:extr_der_Qtilde} $\sim$ \eqref{eq:extr_der_chi}.
The yellow (blue) part represents the success (failure) phase of the reconstruction defined by the logarithm of the MSE.
The results are in good correspondence with the ones computed by ADMM illustrated in Figs.~\ref{fig:phase_deviation}(b) and (d).
Since the replica analysis assumes an infinite system, 
whereas ADMM is only applicable to a finite system; the results show $N=1000$,
the diagrams do not completely match, and the replica analysis shows the MEM reconstruction limit.

When we compare $\rho = 0.8$ with $\rho = 0.2$, the MSE is improved because the number of "lie" included in the $\rho = 0.8$ model is smaller than  $\rho = 0.2$.
Since in the deviation model, we consider the deviation for only $X^0_i = 0$ as $d_i = \varepsilon$,
the more the sparsity is, the smaller the deviation is.
Thus the high value of $\rho$ system can be reconstructed better.

We analyze the nature of the phase transition.
The transition occurs as a function of $\alpha$ even though the MSE is increasing continuously on one of $\varepsilon$ illustrated in Fig.~\ref{fig:transition}(a) and (b).
Figure~\ref{fig:transition}(a) shows  the change in MSE with respect to $\alpha$ for different values of $\rho$.
It is demonstrated that as the sparsity increases ($\rho$ decrease), the phase transition point shifts to the left.
This phenomenon is also attributed to the greater mismatch between the original signal and the default model 
when the sparsity is high, as previously mentioned.

\subsection{Flipping model} \label{sub_res_model1}
\begin{figure}[t]
    \centering
    \includegraphics[width=1.0\linewidth]{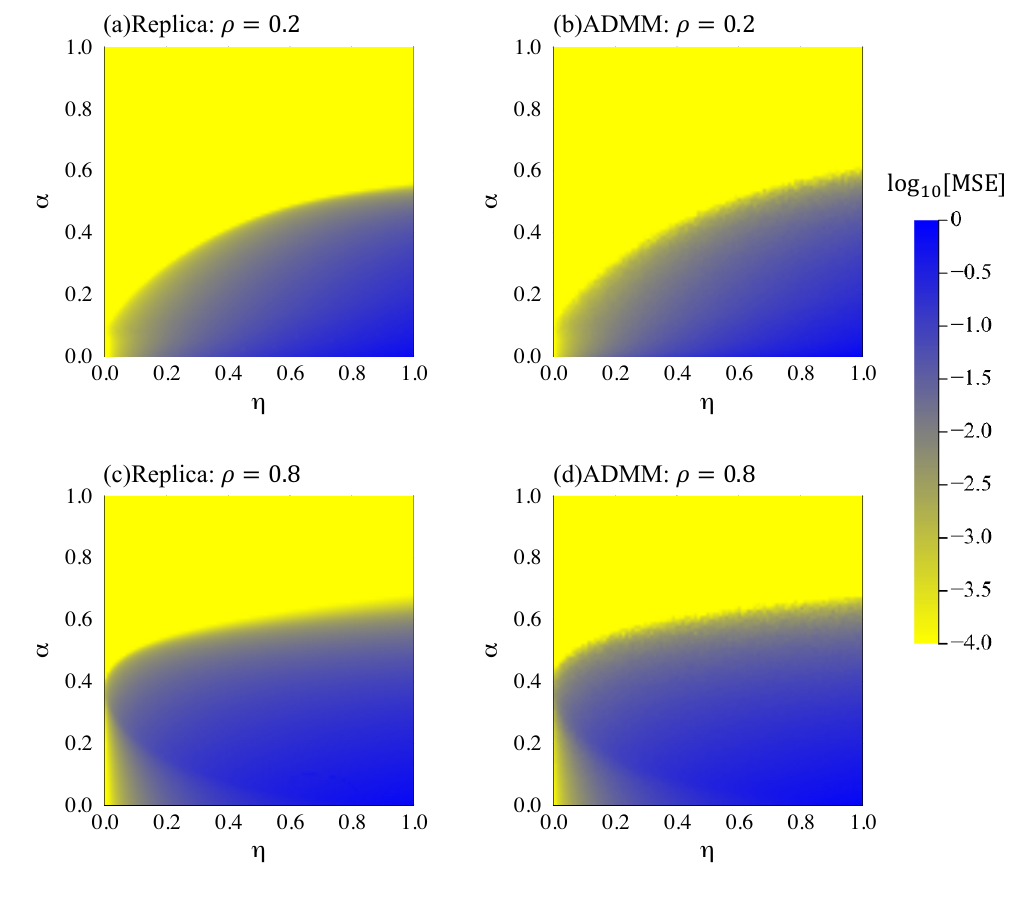}
    \caption{
        The phase diagram of the flipping model. 
        (a)The result of the replica analysis and (b)one of ADMM calculation for $\rho=0.2$.
        (c)The result of the replica analysis and (d)one of ADMM calculation for $\rho=0.8$.
    }
    \label{fig:phase_flipping}
\end{figure}

In the flipping model, 
the phase diagram different from that of the deviation model emerges.
Figure~\ref{fig:phase_flipping}(a) and (c) show the phase diagrams for $\rho = 0.2$ and $0.8$ calculated by the replica method,
and (b) and (d) are the results of ADMM.
The results are in good agreement.

\begin{figure}[t]
    \centering
    \includegraphics[width=0.9\linewidth]{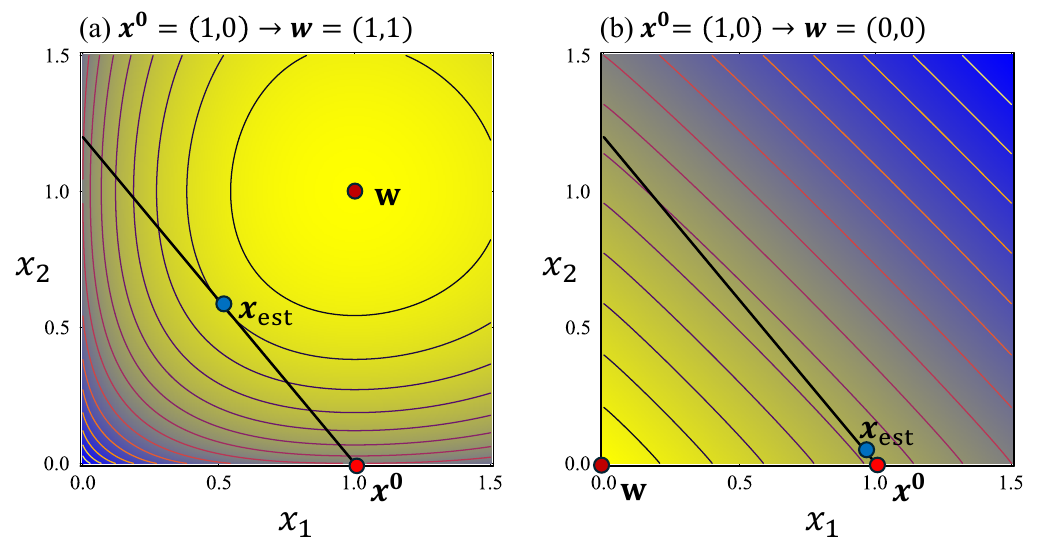}
    \caption{
        Illustration of the impact of flipping in the default model on the estimated signal.
        (a)When the original signal is $\bm{x}^0 = (1,0)$ and the default model perfectly matches it; $\bm{w} = (1,1)$,
        the reconstructed values $\bm{x}_{\text{est}}$ closely align with $\bm{x}^0$.
        (b) When the default model is flipped as \hitomi{$\bm{w} = (0,0)$},
        the estimated values shift significantly, 
        demonstrating the influence of default model discrepancies on MEM reconstruction.
    }
    \label{fig:MEM_N2M1}
\end{figure}

First, we focus on the difference between the phase diagram of $\rho=0.2$ (a) and $0.8$ (c).
In the flipping model, 
$\eta$ represents the flip ratio of the default model relative to the original signal and 
is independent of the sparsity of the original signal $\rho$. 
Nevertheless, 
the phase diagrams for $\rho = \rho_0$ and $\rho = 1-\rho_0$ are generally different, 
indicating that the reconstruction accuracy breaks symmetry with respect to $\rho$.
This asymmetry arises due to the following reason: in the flipping model, 
the estimation results differ depending on whether the default model flips an element of the original signal 
from $0$ to $1$ or from $1$ to $0$.
Figure~\ref{fig:MEM_N2M1}(a) is a schematic picture showing the estimated $x$ when $(N, M) = (2, 1)$,
and $\bm{x}^0 = (1,0)$ with $\bm{w} = (1,1)$.
Compared to the case of $w = (1,1)$ shown in Fig,~\ref{fig:MEM_N2M1}(b), 
it can be observed that both the estimation result obtained by MEM and its accuracy differ. 
Note that the MSE is defined as the squared distance between $\bm{x}_0$ and $\bm{x}$.

\begin{figure}[t]
    \centering
    \includegraphics[width=1.0\linewidth]{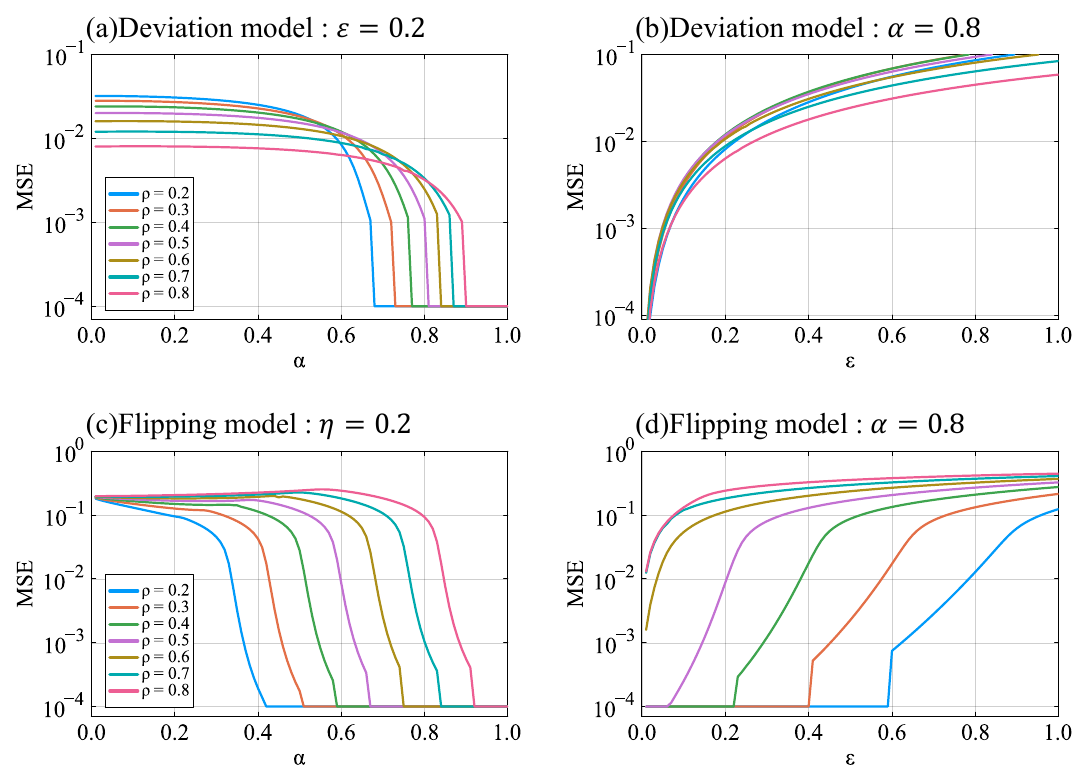}
    \caption{
    The $\rho$-dependence of MSE.
    (a) and (b) show dependece about $\alpha$ and $\varepsilon$ of the deviation model.
    (c) and (b) show dependece about $\alpha$ and $\eta$ of the flipping model.
    Even though there is no transition point at $\varepsilon$ derection of the deviation model,
    the point cause at $\eta$ of the deviation model.
    This is because the flipping model keeps discrete variables at each set up.
    }
    \label{fig:transition}
\end{figure}

Furthermore, 
while no phase transition exists in the direction of the deviation $\varepsilon$ in the deviation model, 
the flipping model exhibits phase transition points in the $\eta$ direction, as shown in Fig.~\ref{fig:transition}(d). 
This can be attributed to the fact that the deviation in the former is continuous, 
whereas in the latter, the default model always takes discrete values of either $0$ or $1$.

\section{Comparison with $L_1$-norm regularization} \label{sec_res_L1}
\begin{figure*}[!htb]
    \centering
    \includegraphics[width=1.0\linewidth]{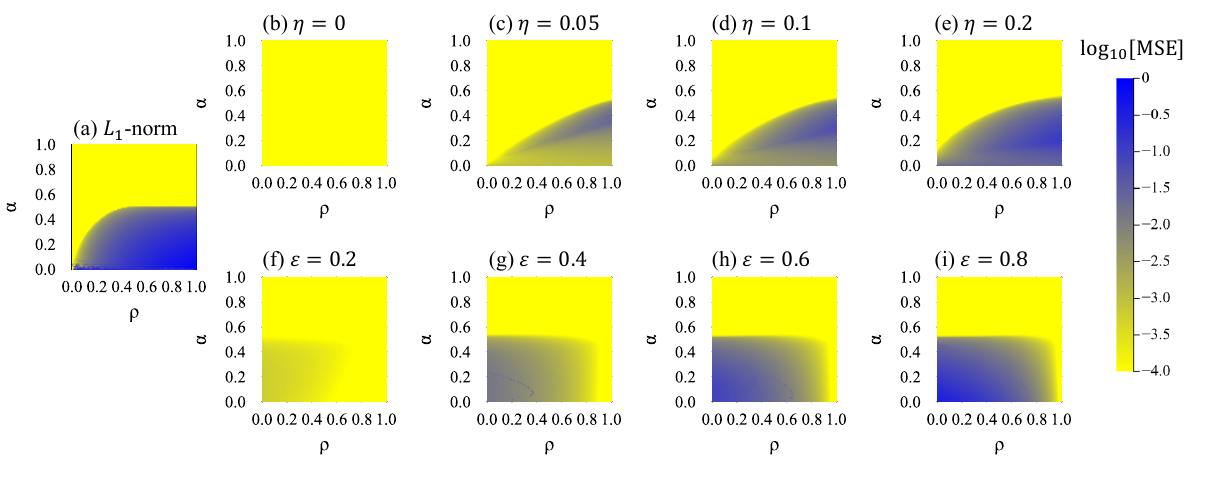}
    \caption{
        \hitomi{
        The phase diagram of
        (a)$L_1$-norm regularization,
        (b) $\sim$ (e)MEM estimation of the flipping model with $\eta = 0$, $0.05$, $0.1$, $0.2$.
        Even though the flipping model with $\eta = 0$ shown in (b) is represent the original signal perfectly,
        the reconstruction with the finite $\eta$ causes the failure phase in (c) $\sim$ (e).
        (f) $\sim$ (i)MEM estimation of the deviation model with $\varepsilon = 0.2$, $0.4$, $0.6$, $0.8$.
        When the original signal is fully dense, i.e. $\\rho = 1$, 
        the original signal and the default model coincide completely, and perfect reconstruction becomes possible.
        The horizontal and vertical axis of every figure denote the sparsity $\\rho$ and $\\alpha$.
        }
    }
    \label{fig:vsL1}
\end{figure*}
$L_1$-norm regularization is one of the methods used for signal reconstruction. 
This method assumes that the original signal is sparse, and does not require a default model. 
\hitomi{
The method can be formulated as 
\begin{equation}
    \underset{\{0 \leq x \leq 1\}^N}{\min} |\bm{x}|_1 \quad \text{subject to} \quad \bm{y} = A \bm{x},
    \label{eq:L1_minimization}
\end{equation}
where $|\bm{x}|_1$ denotes $L_1$-norm of vector $\bm{x}$.
While MEM reconstruction does not requre sparsity of original signals,
the method is considered effective for reconstructing signals when the original signal is sparse.
}
In this section,
we compared the reconstruction performance of the flipping model and $L_1$-norm regularization by the replica approach.

Figure~\ref{fig:vsL1} shows the phase diagram of (a)$L_1$-norm regularization, (b) $\sim$ (e)MEM estimation with $\eta = 0$, $0.05$, $0.1$, and $0.2$.
The result of $L_1$-norm regularization is correspond to the previous study
\hitomi{
    \cite{doi2024phase}
}.
In Fig.~\ref{fig:vsL1}(b), 
the default model perfectly matches the original signal, 
resulting in successful reconstruction in all regions. 
On the other hand, 
\hitomi{
    in Fig.~\ref{fig:vsL1}(c), for example, 
    Even with only a $5\%$ deviation of the default model from the original signal,
    a distinct failure region is observed.
    Fig.~\ref{fig:vsL1}(d) and (e) show the cases with $10\%$ and $20\%$ deviations, 
    where the MSE in the reconstruction failure region at large $\rho$ becomes comparable to that of the $L_1$-norm method.
}
This result prominently reflects the dependence of the MEM on the default model, 
indicating that when the default model deviates even slightly from the original signal, 
the failure region arises in the reconstruction. 
This is due to the phase transition mentioned in the previous section, 
which occurs even for small values of $\eta$ representing a critical weakness in the reconstruction process using MEM.

Moreover, in general, 
$L_1$-norm regularization leverages the sparsity of the original signal. 
It is considered that MEM estimation, which does not rely on such an assumption, 
would be more effective in regions with low sparsity (regions where $\rho$ is large). 
However, the results in Fig.~\ref{fig:vsL1} indicate that 
even in dense regions of the original signal, $L_1$-norm regularization and MEM achieve comparable levels of reconstruction accuracy.

\hitomi{
    In the case of the deviation model, 
    as discussed in Sec.~\ref{sub_model1}, 
    the phase diagram is significantly different from that of the flipping model.
    We illustrated it in Fig.~\ref{fig:vsL1}(f) $\sim$ (i).
    In the deviation model, 
    the discrepancy $\varepsilon$ of the default model is taken into account only when the value of the original signal is zero. 
    Therefore, 
    when the original signal is fully dense, i.e. $\rho = 1$, 
    the original signal and the default model coincide completely, and perfect reconstruction becomes possible as shown higher $\rho$ reagion in Fig.~\ref{fig:vsL1}(f) $\sim$ (i). 
    In contrast, 
    when $\rho$ is small, i.e., when the original signal is sparse, the relative discrepancy becomes larger, resulting in an expansion of the reconstruction failure region. 
}

\section{Conclusion} \label{sec_con}
In this study, 
we clarified the dependence of MEM reconstruction accuracy on the default model
\hitomi{
    for binary signals
}. 
The results revealed that discrepancies between the default model and the original signal lead to the emergence of success and failure phases in reconstruction. 
Specifically, in the deviation model, 
where discrepancies are continuous, phase transition points were observed only in the $\alpha$ direction. 
In contrast, in the flipping model, 
which assumes a default model obtained by flipping the original signal, 
phase transitions occurred not only in the $\alpha$ direction but also in the $\eta$ direction, 
corresponding to the flip ratio. 
This indicates that reconstruction becomes impossible if a finite discrepancy exists, 
even if the discrepancy is small.

Furthermore, 
a comparison with $L_1$-norm regularization, 
which has recently been proposed as a new reconstruction method, 
revealed that MEM achieves reconstruction accuracy equivalent to or lower than that of $L_1$-norm regularization, 
not only in sparse regions where $L_1$-norm regularization excels but also in dense regions.

While MEM is a classical and versatile reconstruction method widely used across various scientific fields, 
our findings suggest that caution should be exercised when applying it.
If the assumed default model deviates from the true distribution, 
reliable results cannot be obtained. 
Future research should evaluate MEM’s applicability and default models in various domains.

\begin{acknowledgments}
    MH thank the fruitful discussions with Mikiya Doi. 
    This work is supported by JSPS KAKENHI Grant No.23H01432.
    Our study receives financial support from the programs 
    for Bridging the gap beween R\&D and the IDeal society (society 5.0) and Generating Economic and social value (BRIDGE)
    and the Cross-ministerial Strategic Innovation Promotion Program (SIP) from the Cabinet Office.
\end{acknowledgments}

\appendix
\begin{widetext}
\section{Derivation of free energy density}
We presents the detailed derivation of the free energy density Eq.~\eqref{eq:extr_free_energy_density}.
Starting with Eq.~\eqref{eq:replicated_system}, it is possible to be rewritten by the order parametes Eqs.~\eqref{eq:orders_rho} $\sim$ \eqref{eq:orders_q} as follows :
\begin{align}
    \left[Z_{\beta}^{n}(A, \bm{y})\right]_{A, \bm{x}^{0}, \bm{w}}
    =
    \prod_{a}
    \int_{0}^{1} d \bm{x}^{a} 
    \int d \bm{x}^{0}
    P(\bm{x}^{0})
    \int d \bm{w}
    P(\bm{w})
    \exp(-\beta S(\bm{x}^{a}, \bm{w}))
    \prod_{\mu=1}^{M}
    \left[
        \prod_{a}
        \delta(u_{\mu}^{a}) 
    \right]_{\xi_{a}, z} 
    \nonumber \\
    \times
    \prod_{a, b}
    \int{dm} \int{dQ} \int{dq} \,
    \delta \left(m - \frac{1}{N}{\bm{x}^{0}}^{T} \bm{x}^{a} \right)
    \delta \left(Q - \frac{1}{N}{\bm{x}^{a}}^{T} \bm{x}^{a} \right)
    \delta \left(q - \frac{1}{N}{\bm{x}^{a}}^{T} \bm{x}^{b} \right)
    \label{eq:Z_with_order}
\end{align}
where $P(\bm{w})$ is the probability distribution of the default models Eq.~\eqref{eq:der_distribution} and Eq.~\eqref{eq:flp_distribution}.
We introduce the conjugate variables of the order parameters Eqs.~\eqref{eq:orders_rho} $\sim$ \eqref{eq:orders_q} by Fourier transformation representation of $\delta$ function,
\begin{align}
    \delta \left(m - \frac{1}{N}{\bm{x}^{0}}^{T} \bm{x}^{a} \right)
    & =
    \int d \tilde{m} \exp\left(- \tilde{m}\left(Nm - {\bm{x}^{0}}^{T} \bm{x}^{a} \right) \right),
    \label{eq:delta_integral_m}
    \\
    \delta \left(Q - \frac{1}{N}{\bm{x}^{a}}^{T} \bm{x}^{a} \right)
    & = 
    \int d \tilde{Q} \exp\left(\frac{\tilde{Q}}{2}\left(NQ - {\bm{x}^{a}}^{T} \bm{x}^{a} \right) \right),
    \label{eq:delta_integral_Q}
    \\
    \delta \left(q - \frac{1}{N}{\bm{x}^{a}}^{T} \bm{x}^{b} \right) 
    & =
    \int d \tilde{q} \exp\left(- \frac{\tilde{q}}{2}\left(Nq - {\bm{x}^{a}}^{T} \bm{x}^{b} \right) \right).
    \label{eq:delta_integral_q}
\end{align}
We also apply the Hubbard-Stratonovich transformation,
\begin{align}
    \prod_{a\neq b}\exp\left(\frac{\tilde{q}}{2}{\bm{x}^{a}}^{T} \bm{x}^{b} \right)
    & =
    \int Dt \, 
    \prod_{a = 1}^{n}
    \prod_{i=1}^{N}
    \exp\left(\sqrt{\tilde{q}} x^{a}_{i} t - \frac{\tilde{q}}{2} (x^{a}_{i})^{2}\right),
    \label{eq:habast}
\end{align}
to Eq.~\eqref{eq:delta_integral_q}, we represent Eq.~\eqref{eq:Z_with_order} as
\begin{align}
    \left[Z_{\beta}^{n}(A, \bm{y})\right]_{A, \bm{x}^{0}, \bm{w}} 
    & =
    \int dm \int dQ \int dq 
    \int d\tilde{m} \int d\tilde{Q} \int d\tilde{q} \nonumber
    \\
    & \times
    \exp
    \left(nN 
    \left(
        - \frac{\alpha}{2(Q - q)}(\rho - 2m + q)
        - \frac{\alpha}{2} \log(2\pi)
        - m \tilde{m} + \frac{1}{2}Q \tilde{Q} + \frac{1}{2}q \tilde{q} 
    \right) 
    \right) 
    \nonumber \\ 
    & \times
        \left[
        \int Dt \,
            \prod_{a=1}^{n}
            \prod_{i=1}^{N}
            \left(
                \int_{0}^{1} dx^{a}_{i}
                \exp 
                \left(
                    - \frac{1}{2}(\tilde{Q} + \tilde{q})(x^{a}_{i})^{2} 
                    + (\tilde{m}x^{0}_{i} + \sqrt{\tilde{q}} t)x^{a}_{i} 
                    + \beta s(x^a ; w_i)
                \right)
            \right)
        \right]_{x^{0}_{i}, \bm{w}}.
        \label{eq:replicated_system_explicit}
\end{align}
Since we cosider the limit $\beta \rightarrow \infty$, 
the order parameters satisfy the saddle points approximation, $\int dx \exp(Ng(x)) \approx \exp(Ng(x^{*}))$ for the integrals.
Equation~\eqref{eq:replicated_system_explicit} can be rewritten as
\begin{align}
    \left[Z_{\beta}^{n}(A, \bm{y})\right]_{A, \bm{x}^{0}} 
    \approx
    \exp\left(nN 
    \left(
        - \frac{\alpha}{2(Q - q)}(\rho - 2m + q)
        - m \tilde{m} + \frac{1}{2}Q \tilde{Q} + \frac{1}{2}q \tilde{q} 
    \right) 
    \right) 
    \nonumber \\
    \times
        \left[
            \int Dt \,
            \exp 
            \left(nN
            \log
            \left(
                \int_{0}^{1}dx
                \exp 
                \left(
                    - \frac{1}{2}(\tilde{Q} + \tilde{q})x^{2} 
                    + (\tilde{m}x^{0} + \sqrt{\tilde{q}} t + \tilde{p})x 
                    + \beta s(x; w)
                \right)
            \right)
            \right)
        \right]_{x^{0}, w} 
    \label{eq:replicated_system_explicit2}
\end{align}
where the saddle point of every order parameter is denoted as $x^{*} \rightarrow x$ (ex. the saddle point $Q^{*}$ as just $Q$),
and the replica and vector indeces are ommited.
We assume that $Q - q \rightarrow \chi/\beta, \tilde{p} \rightarrow \beta \tilde{p}, \tilde{m} \rightarrow \beta \tilde{m}, \tilde{Q} + \tilde{q} \rightarrow \beta \tilde{Q}, \tilde{q} \rightarrow \beta^{2} \tilde{\chi}$, 
as the effect of the templeture.
With the assumptin, a part of Eq.~\eqref{eq:replicated_system_explicit2} is expressed as follow,
\begin{align}
    &\left[
        \int Dt
        \exp 
        \left(nN
        \log
        \left(
            \int_{0}^{1}dx
            \exp 
            \left(
                - \frac{1}{2}(\tilde{Q} + \tilde{q})x^{2} 
                + (\tilde{m}x^{0} + \sqrt{\tilde{q}} t)x 
                + \beta s(x; w)
            \right)
        \right)
        \right)
    \right]_{x^{0},w}
    \nonumber \\
    & \rightarrow
    \left[
        \int Dt
        \exp 
        \left(nN
        \log
        \left(
            \int_{0}^{1}dx
            \exp
            \left(
            \beta
                \left(
                    - \frac{1}{2}\tilde{Q}x^{2} 
                    + (\tilde{m}x^{0} + \sqrt{\tilde{\chi}} t + \tilde{p})x 
                    + s(x; w)
                \right)
            \right)
        \right)
        \right)
    \right]_{x^{0},w} 
    \nonumber \\
    & \approx
    \exp
    \left(
        nN
        \left[
        \int Dt \,
            \log
            \left(
                \int_{0}^{1}dx
                \exp
                \left(
                \beta
                \left(
                    - \frac{1}{2}\tilde{Q}x^{2} 
                    + (\tilde{m}x^{0} + \sqrt{\tilde{\chi}} t)x 
                    + s(x; w)
                \right)
                \right)
            \right)
        \right]_{x^{0},w}
    \right) ~~ (\because n \rightarrow 0)
    \nonumber \\
    & \approx
    \exp
    \left(
        nN\beta
        \left[
            \int Dt \,
            \max_{\{0 \leq x \leq 1\}} 
            \left\{
                -\frac{1}{2} \tilde{Q}x^{2} 
                + \left(\tilde{m} x^{0} + \sqrt{\tilde{\chi}} t \right)x
                + s(x; w)
            \right\}
        \right]_{x^{0},w}
    \right) ~~ (\because \beta \rightarrow \infty) 
    \nonumber \\
    & =
    \exp
    \left(
        -nN\beta
        \left[
            \int Dt \,
            \Psi (t,x^{0},w; \tilde{Q},\tilde{m},\tilde{\chi})
        \right]_{x^{0},w}
    \right).
    \label{eq:part_explicit}
\end{align}
Finally, we obtain Eq.~\eqref{eq:extr_free_energy_density}.

\section{Derivation of Eq.\eqref{eq:x_ast}}
Equation~\eqref{eq:x_ast} is derived as bellow:

$x^{*}$ is the minimum value of the equation,
\begin{align}
    \psi (x; t,x^{0},w, \tilde{Q},\tilde{m},\tilde{\chi})
    & = \frac{\tilde{Q}}{2} x^2 - (\tilde{m}x^0 + \sqrt{\tilde{\chi}}t)x - s(x; w) \\
    & = \frac{\tilde{Q}}{2} x^2 - (\tilde{m}x^0 + \sqrt{\tilde{\chi}}t)x - x + w + x \log \frac{x}{w}.
\end{align}
Since the function is convex,
the minimum value can be derived by computing the zero point of the differential coefficient.
\begin{align}
    \partial_{x} \psi (x; t,x^{0},w, \tilde{Q},\tilde{m},\tilde{\chi}) &= 0 \nonumber \\
    \tilde{Q} x - (\tilde{m}x^0 + \sqrt{\tilde{\chi}}t) + \log x - \log w &= 0 \nonumber \\
    \log \left( \tilde{Q} x \mathrm{e}^{\tilde{Q} x} \right) &= \log \left(\tilde{Q} w \mathrm{e}^{(\tilde{m}x^0 + \sqrt{\tilde{\chi}}t)} \right) \nonumber \\
    \tilde{Q} x \mathrm{e}^{\tilde{Q} x} &= \tilde{Q} w \mathrm{e}^{(\tilde{m}x^0 + \sqrt{\tilde{\chi}}t)} \nonumber \\
    \tilde{Q} x &= W \left(\tilde{Q} w \mathrm{e}^{(\tilde{m}x^0 + \sqrt{\tilde{\chi}}t)} \right) \nonumber \\
    x &= \frac{1}{\tilde{Q}} W \left( \tilde{Q} w \mathrm{e}^{(\tilde{m}x^0 + \sqrt{\tilde{\chi}}t)} \right),
\end{align}
where $W(z)$ is Lambert W-function satisfying
\begin{equation}
    W(z) \mathrm{e}^{W(z)} = z.
\end{equation}

\end{widetext}

\bibliography{references}

\end{document}